\documentclass[aps,amsmath,twocolumn,a4paper,pre,showpacs]{revtex4}

\usepackage{graphicx}
\usepackage{amssymb}

\newcommand {\bee} {\begin{equation}}
\newcommand {\eee} {\end{equation}}
\newcommand {\bea} {\begin{eqnarray}}
\newcommand {\eea} {\end{eqnarray}}
\newcommand {\bes} {\begin{displaymath}}
\newcommand {\ees} {\end{displaymath}}
\newcommand {\beas} {\begin{eqnarray*}}
\newcommand {\eeas} {\end{eqnarray*}}
\newcommand {\comment}[1]{}
\newcommand {\ctwo} {{HT}}
\newcommand {\cone} {{LT}}
\newcommand {\drho} {\rho}
\newcommand {\prho} {\varrho}

\newcommand{\llangle}{<}
\newcommand{\rrangle}{>}
\begin{document}

\title{Universality and non-universality of mobility in heterogeneous
  single-file systems and Rouse chains}

\author{Michael A. Lomholt}
\email{mlomholt@memphys.sdu.dk}
\affiliation{MEMPHYS - Center for Biomembrane Physics, Department of Physics, Chemistry and Pharmacy,
University of Southern Denmark, Campusvej 55, 5230 Odense M, Denmark}

\author{Tobias Ambj\"ornsson}
\affiliation{Department of Astronomy and Theoretical Physics, Lund University,
  S\"olvegatan 14A, SE-223 62 Lund, Sweden}

%%%%%%%%%%%%%%%%%%%%%%%%%%%%%%%%%%%%%%%%%%%%%%%%%%%%%%%%%%%%%%%%%%%%%%%%%%%%%%%%
%
%	A b s t r a c t / I n t r o d u c t o r y   p a r a g r a p h
%
%%%%%%%%%%%%%%%%%%%%%%%%%%%%%%%%%%%%%%%%%%%%%%%%%%%%%%%%%%%%%%%%%%%%%%%%%%%%%%%%

\begin{abstract}
  
  We study analytically the tracer particle mobility in single-file
  systems with distributed friction constants.  Our system serves as a
  prototype for non-equilibrium, heterogeneous, strongly interacting
  Brownian systems. The long time dynamics for such a single-file
  setup belongs to the same universality class as the Rouse model with
  dissimilar beads. The friction constants are drawn from a density
  $\prho(\xi)$ and we derive an asymptotically exact solution for the
  mobility distribution $P[\mu_0(s)]$, where $\mu_0(s)$ is the
  Laplace-space mobility.  If $\prho$ is light-tailed (first moment
  exists) we find a self-averaging behaviour:
  $P[\mu_0(s)]=\delta[\mu_0(s)-\mu(s)]$ with $\mu(s)\propto
  s^{1/2}$. When $\prho(\xi)$ is heavy-tailed, $\prho(\xi)\simeq
  \xi^{-1-\alpha} \ (0<\alpha<1)$ for large $\xi$ we obtain moments
  $\langle [\mu_s(0)]^n\rangle \propto s^{\beta n}$ where
  $\beta=1/(1+\alpha)$ and no self-averaging.  The results are
  corroborated by simulations.

\end{abstract}


\maketitle

%%%%%%%%%%%%%%%%%%%%%%%%%%%%%%%%%%%%%%%%%%%%%%%%%%%%%%%%%%%%%%%%%%%%%%%%%%%%%%%%
%
%			M a i n   t e x t
%
%%%%%%%%%%%%%%%%%%%%%%%%%%%%%%%%%%%%%%%%%%%%%%%%%%%%%%%%%%%%%%%%%%%%%%%%%%%%%%%%

\section{Introduction}

Studies of the force response properties in complex media have a long tradition
in physics \cite{kubo_1986,marconi_2008}. In biology, forces are involved in a
large number of different processes in cells, and moreover, forces are
commonly used in force probing, for instance, of macromolecular structure in
in vitro systems \cite{bustamante_04}. The Jarzynski equality relates the
time-averaged response of a system when under influence of a force to the free
energy between initial and final states \cite{jarzynski}. Recently, an exact
solution to a paradigm non-equilibrium model for homogeneous systems,
the asymmetric exclusion process, was put forward \cite{gorissen2012}.

In this article we provide asymptotically exact solutions for the force
response of a complex {\em heterogeneous} system: tracer particle dynamics in
a single-file system (same universality class as harmonically coupled
dissimilar beads or Rouse chains, for long times) with randomly distributed friction constants.
Our model serves as a prototype for the non-equilibrium dynamics in
heterogeneous, strongly interacting Brownian systems. Even for the case when
all particles have identical friction constants such systems display
non-trivial dynamics characterized by a subdiffusive behaviour
\cite{HA,Kollmann_03,Taloni_06,barkai_09}. Fewer studies have addressed the
problem of diffusion of hardcore particles with different friction constants,
for undriven systems see
Refs. \cite{Aslangul_00,Brzank2,JALA,Jara_09,TA_etal,Flomenbom,em_sfd}.  Of
particular interest for the present study is Ref. \cite{em_sfd} where an
effective medium approximation was applied revealing ultra-slow time-evolution
of the mean square displacement and simulations indicated lack of
self-averaging. To our knowledge the problem addressed in this paper, namely
the exact force-response relation for tracer particle dynamics in single-file
systems with distributed friction constants, has not been addressed
previously. From our treatment of these systems we also obtain exact results for the mean square displacement of the tracer particle.

Besides its theoretically interesting properties, the single-file problem
finds a number of experimental realizations: transport in microporous
materials~\cite{KKDGPRSUK, meersmann00,hahn96} (e.g. zeolites), colloidal
systems~\cite{WBL}, molecular sieves \cite{gupta95} and biological
pores~\cite{HOKE}. Cooperative effects are of importance in transport
processes involving molecular motors \cite{vermeulen_2002,berger_2012}.
Hardcore repulsion of binding proteins diffusing along DNA has been shown to
be important in transcription~\cite{Elf_09}.

\section{Description of the system}

\begin{figure}[tb]
\begin{center}
\includegraphics[width=0.8\columnwidth]{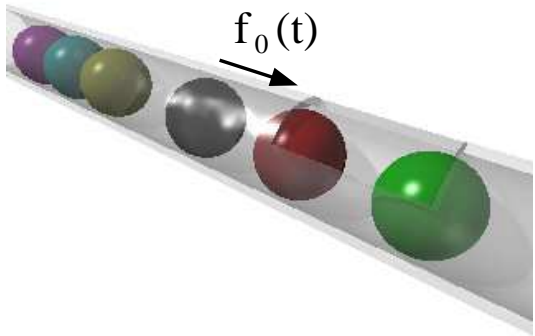}
\end{center}
\caption{Cartoon of the heterogeneous single-file system investigated in this
  article. Dissimilar hardcore interacting particles (the particles cannot
  overtake) are diffusing in a one-dimensional system. The particles are
  assigned {\em different} friction constants, $\xi_n$ ($n$ labels different
  beads), drawn from a probability density $\prho(\xi_n)$. A time-varying
  force, $f_0(t)$, acts on a tracer particle (colored black). In such a scenario we study the
  tracer particle force response properties, through the mobility defined in
  Eq. (\ref{eq:mu_0}).}\label{fig:cartoon}
\end{figure}

Let us state the problem. We consider strongly overdamped motion of Brownian
particles, in an infinite one dimensional system, interacting via a
two-body short-range repulsive potential. This
potential, ${\cal V}(|x_n(t) - x_{n'}(t)|)$, where $x_n(t)$ is the
position of the $n$th particle, has a hard-core part which excludes
particles from overtaking each other.  The Langevin equations of
motion are thus
$\xi_n \dot{x}_n(t) = \sum_{n'}
\mathfrak{f}[x_n(t) - x_{n'}(t)] +\eta_n (t) + f_0 (t)\delta_{n,0}$
where a dot denotes time derivative, $\mathfrak{f}=-\partial {\cal
  V}/ \partial x_n$ is the interaction force, $\eta_n(t)$ is a Gaussian zero-mean noise,
$\llangle\eta_n(t)\rrangle =0$, with correlations that are related to the friction constants $\xi_n$ by the
fluctuation-dissipation theorem \cite{kubo66} to be
$\llangle\eta_n(t)\eta_{n'}(t')\rrangle =2k_BT\xi_n\delta(t-t')\delta_{n,n'}$,
where $k_B$ is the Boltzmann constant and $T$ the temperature.  $f_0(t)$ is an
external force acting only on particle $0$ (the tracer particle). 
In our simulations we take
$f_0(t)$ to be an oscillating force. A cartoon of the problem at hand is
depicted in Fig. \ref{fig:cartoon}.

As was shown in \cite{lizana_10,em_sfd} using a {\it harmonization} approach
 the long time limit of the Langevin equation above with a sufficiently small external force $f_0(t)$ is the same as that for a linear chain of interconnected springs
\bea
\xi_n\frac{d x_n(t)}{d t}&=&\kappa\left[x_{n+1}(t)+x_{n-1}(t)-2
  x_n(t)\right]+\eta_n(t)\nonumber\\
&&+f_0(t)\delta_{n,0}\label{eq:eq_of_motion},
\eea
The effective nearest neighbor
spring constant $\kappa$ is obtained from the system's equation of state. For
hardcore interacting particles of size $b$ (used in our simulations) this
harmonization procedure yields \cite{lizana_10} $\kappa = \drho^2 k_B T (1 -
\drho b )^{-2}$, where $\drho$ is the particle density.

The heterogeneity of the particles enters through their different friction
constants, $\xi_n$, which here are assumed to be identically distributed
random variables taken from a probability density $\prho(\xi_n)$. We
distinguish between light-tailed ({\cone}) distributions for which the mean
$\bar{\xi}$ of $\prho(\xi)$ exists, and heavy-tailed ({\ctwo}) systems, where
\bee\label{eq:prho}
\prho(\xi)\sim A\xi^{-1-\alpha}
\eee 
for large $\xi$, with $A$ a constant prefactor and $0<\alpha<1$ such that the mean diverges.

The main quantity of interest in this study is the distribution $P[\hat{\mu}_0(s)]$
of mobilities of the tracer particle 0 defined as (in the Laplace domain):
\bee\label{eq:mu_0}
\hat{\mu}_0(s)\equiv \frac{\llangle \hat{v}_0(s)\rrangle}{\hat{f}_0(s)}
\eee
where $v_0(t)=d x_0(t)/dt$ is the tracer particle velocity, and we use a `hat'
to distinguish quantities in Laplace-space [Laplace-transforms are defined
$\hat{A}(s)=\int_0^\infty dt\, e^{-st} A(t)$]. 
Brackets $\llangle .. \rrangle$ represent an average
over different
realizations of the thermal noise and random initial positions. We label this
average the {\em non-averaged} case.
It is contrasted by the {\em heterogeneity-averaged} case (represented by
$\langle .. \rangle$) where an additional average over the probability density
of friction constants is performed. 
%
%Brackets $\langle .. \rangle$ represent an implicit average
%over quenched friction constants, besides averages over different
%realizations of the noise and random initial positions. We label this
%average the {\em heterogeneity-averaged} case \cite{em_sfd},
%which is contrasted by the {\em non-averaged} case (represented by
%$\llangle .. \rrangle$) where an average over the probability density
%of friction constants is not performed. 
%
In the simulations for the
non-averaged case the same $\xi_n$'s are used when averaging over
thermal noise (i.e., for each simulation run). For the heterogeneity
averaged case we draw new friction constants whenever we make a new
initial particle positioning.

Turning back to Eq.  (\ref{eq:eq_of_motion}), introducing the quantity
$y_n(t)=x_n(t)-n/\drho$, and taking the Laplace-transform we obtain
\bea
&&\xi_n [s
\llangle \hat{y}_n(s)\rrangle -\llangle y_n(0)\rrangle
]\nonumber\\
&=&\kappa\left[\llangle\hat{y}_{n+1}(s)\rrangle+\llangle
  \hat{y}_{n-1}(s)\rrangle -2
  \llangle \hat{y}_n(s)\rrangle \right]\nonumber\\
&&+\hat{f}_0(s)\delta_{n,0}\label{eq:eq_of_motion_s}.
\eea
We proceed by introducing the quantities $m_n^{(\pm)}$, representing mobility
of the chain to the right $(+)$ or left $(-)$ starting from particle $n$
(using $\llangle y_n(0)\rrangle=0$), defined as
\begin{equation}
m_n^{(\pm)}(s)=\frac{s \llangle \hat{y}_n(s)\rrangle}{-\kappa(\llangle\hat{y}_n(s)\rrangle-\llangle\hat{y}_{n\mp 1}(s)\rrangle)}
\end{equation}
Notice that the denominator represents the velocity of particle $n$, while the numerator represents the force from one of its harmonic springs. With these definitions we obtain the following expression for the tracer particle mobility of particle 0 \cite{footnote1}:
\begin{equation}
\hat{\mu}_0(s)=\left(\xi_0 +\frac{1}{s/\kappa + m_1^{(+)}(s)}+\frac{1}{s/\kappa
  + m_{-1}^{(-)}(s)}\right)^{-1}\label{eq:mu_0_s}
\end{equation}
as well as the following recurrence relations
\bee\label{eq:recursion1}
m_{\pm n}^{(\pm)}(s)=\left( \xi_n + \frac{1}{s/\kappa + m_{\pm (n+ 1)}^{(\pm)}(s)}\right)^{-1},\; n>0.
\eee
For a given set of $\xi_n$'s Eqs. (\ref{eq:mu_0_s}) and
(\ref{eq:recursion1}) provide an exact expression for the tracer particle 
mobility.

\section{Tracer mobility for $\xi_n$ being iid random variables}

We now proceed by assuming that the $\xi_n$'s are independent,
identically distributed (iid) random numbers, and try to solve
Eqs. (\ref{eq:mu_0_s}) and (\ref{eq:recursion1}) for the probability distribution of $\hat{\mu}_0(s)$. Note that since
the $\xi_n$'s are identically distributed random variables, so are the
$m_n$'s; we denote by $g_s (m_n)$ the corresponding distribution. We
obtain an equation for $g_s (m)$ by writing down the formula for the distribution of $m_n^{(\pm)}$ in terms of the identical distribution of $m_{\pm(n+1)}^{(\pm)}$. By Eq. (\ref{eq:recursion1}) it is
\bea\label{eq:g_s_eq}
g_s(m)&=&\int_0^\infty dm' g_s(m')\int_0^\infty dy\, R(y)\nonumber\\
&&\times \delta
\left( m-\left(\frac{1}{y} +\frac{1}{s/\kappa +m'}\right)^{-1}\right),
\eea
where we made the variable substitution $y=1/\xi$ with $R(y)=\varrho(1/y)/y^2$ denoting the corresponding distribution. The function
$\delta(z)$ is the Dirac delta-function. Eq. (\ref{eq:g_s_eq}) constitute an
integral equation for $g_s (m)$.

The probability density for the mobility
(in Laplace-space) is obtained by integrating over all $m$'s
and $y$'s consistent with Eq. (\ref{eq:mu_0_s}):
\begin{align}\label{eq:P_mu0}
&P[\hat{\mu}_0(s)]=\int_0^\infty dy\, R(y)\int_0^\infty dm\, g_s(m)\int_0^\infty dm' g_s(m')\nonumber\\ 
&\;\;\times \delta \left(\hat{\mu}_0(s) -\left( \frac{1}{y}+ \frac{1}{s/\kappa + m}+\frac{1}{s/\kappa  + m'}\right)^{-1}\right).
\end{align}
Eqs. (\ref{eq:g_s_eq}) and (\ref{eq:P_mu0}) define the problem to be
solved. In the following we give asymptotically exact results for the limit
$s\to 0$ (long times).

\subsection{{\cone} systems}
 
Let us first give the results for the quantity of interest, i.e. the
tracer particle mobility probability density Eq. (\ref{eq:P_mu0}), for
{\cone} systems. We make use of the explicit expression for $g_s(m)$
contained in Eqs. (\ref{eq:g_w_scaling}), (\ref{eq:eps_C1}) and
(\ref{eq:h_C1}) in the appendix and find, for $s\rightarrow 0$, that:
\bee\label{eq:P_LT}
P[\hat{\mu}_0(s)]=\delta \big{(} \hat{\mu}_0(s) - \hat{\mu}_0(s)|_{\rm EM,LT} \big{)}
\eee
where  
\bee\label{eq:mu_EM1}
\hat{\mu}_0(s)|_{\rm EM,LT}\sim \frac{s^{1/2}}{2(\kappa\bar{\xi})^{1/2}}
\eee
From Eqs. (\ref{eq:P_LT}) and (\ref{eq:mu_EM1}) we see that {\cone}
systems behave universally at long times like a system of identical
particles all having the friction constant equal to the mean ${\bar \xi}$.  

The result for the tracer particle mobility contained in Eq. (\ref{eq:mu_EM1})
is identical to the effective medium  mobility obtained in
\cite{em_sfd} (appendix A) for {\cone} systems. This effective medium
approximation consists of replacing the disordered quantity $\xi_n$ with a
$n$-independent but instead time-dependent friction kernel $\xi_{\rm eff}(t)$
in such a way that the mobility of a particle on average is unchanged if its
effective friction $\xi_{\rm eff}(t)$ is replaced by one of the original
$\xi_n$. This procedure is thus exact for {\cone} systems at long times.

\subsection{{\ctwo} systems}
 
For the case of {\ctwo} systems, i.e. friction constants drawn from
a distribution with a heavy power-law tail as described by Eq. (\ref{eq:prho}), the analysis is more challenging. As for {\cone}
systems, the problem is divided into two steps, namely, first solve
Eq. (\ref{eq:g_s_eq}) and, second, use the corresponding solution for
$g_s(m)$ to evaluate Eq. (\ref{eq:P_mu0}).

Considering the first step above, we note that if we choose the
specific  type of power-law probability density, $R(y)=\alpha
y^{\alpha-1} $ for $0<y<1$ with $y=1/\xi$ and $R(y)=0$ otherwise,
Eq. (\ref{eq:g_s_eq}) can be solved following the approach in the
appendix of Ref. \cite{bernasconi_80} for long times, $s\rightarrow 0$
(see also Ref. \cite{alexander81}). In Appendix \ref{appA} we
generalize, and simplify, the derivation in \cite{bernasconi_80} to
friction constant probability densities of general type with an
asymptotic behaviour as in Eq. (\ref{eq:prho}).

Let us now turn to the second step, i.e., evaluating
Eq. (\ref{eq:P_mu0}) using the explicit result for $g_s(m)$ obtained in Appendix \ref{appA}.
% as
%provided by Eqs. (\ref{eq:g_w_scaling}), (\ref{eq:eps_C2}) and
%(\ref{eq:h_bernasconi}).
In the limit of $s\rightarrow 0$
Eq. (\ref{eq:P_mu0}) becomes (after a rescaling of the integration
variable):
\bea
P[\hat{\mu}_0(s)]&=&\int_0^\infty dp\, h(p)\int_0^\infty dp'
h(p')\nonumber\\
&&\times \delta\big{(} \hat{\mu}_0(s)-\frac{\epsilon(s)}{1/p+1/p'} \big{)}\label{eq:Pmuexp}
\eea
where the scaling functions $\epsilon(s)$ and $h(q)$ are related to $g_s(m)$ by $g_s(m)=h(m/\epsilon(s))/\epsilon(s)$ with expressions for them provided by Eqs. (\ref{eq:eps_C2}) and (\ref{eq:h_bernasconi}). In arriving at Eq. (\ref{eq:Pmuexp}) we have made use of the normalization condition
$\int_0^\infty R(y) dy=1$. Taking the Mellin-transform
with respect to $\hat{\mu}_0(s)$ of Eq. (\ref{eq:Pmuexp}) we find
\bea
\bar{P}[z]&=&M[P(\mu)]=\int_0^\infty \mu^{z-1} P(\mu) d\mu \\
 &=& [\epsilon(s)]^{z-1} \int_0^\infty dp\, h(p) \int_0^\infty dp'
g(p') f(p'/p)\nonumber
\eea
where  $g(q)=q^{z-1}h(q)$ and $f(q)= (1+q)^{1-z}$.  Using Parseval's
relation for Mellin-transforms, and other standard Mellin-transform relations (see
Ref. \cite{oberhettinger}, Mellin-transform table, Eqs. 1.3 and 2.17)
and interchanging the order of integrations we find
\bee\label{eq:Pintermediate}
\bar{P}(z)=\left(\frac{\beta}{\Gamma(\beta)}\right)^2
\frac{B^{z-1}}{\Gamma(z-1)} \int_{c-i\infty}^{c+i\infty} dw\,
\bar{G}(1-w)\bar{F}(w)
\eee
where $\bar{G}(w)=\Gamma(\beta w)\Gamma(\beta z - \beta w)$ and
$\bar{F}(w)=\Gamma(\beta (z-2)+\beta w)\Gamma(2\beta - \beta w) $
with $B=B(s)=Q(\beta) \mu_0(s)|_{\rm EM,HT}$. We defined the exponent
\bee
\beta=1/(1+\alpha)
\eee
and introduced the result for the mobility within the effective medium approximation for {\ctwo} systems \cite{em_sfd}:
\bee\label{eq:mu_EM2}
{\hat \mu}_0(s)|_{\rm EM,HT}\sim \frac{s^\beta}{2(\kappa\chi)^{1/2}}
\eee
with $ \chi=(4\kappa)^{2\beta-1}
(A\pi/\sin[(1-\beta)\pi/\beta])^{2\beta}$. Also,
\bee
Q(\beta)=\left( \frac{4(1-\beta)}{\beta^3} \Gamma ((1-\beta)/\beta)\right)^\beta.
\eee
In order to arrive at Eq. (\ref{eq:Pintermediate}) we also used
the reflection formula for $\Gamma$-functions
\cite{ABST}.
 Using Parseval's relation in reverse together with standard
Mellin-transforms (see Ref. \cite{oberhettinger}, inverse
Mellin-transform table, Eq. 5.36) we obtain:
$\bar{P}(z)=\big{(}\Gamma(\beta z)/\Gamma(\beta)\big{)}^2
B^{z-1} I/\Gamma(z-1) $ with $I= \int_0^\infty dx
(1+x^{1/\beta})^{-2\beta z}x^{z-2}$. Performing the integral $I$
we get our final expression for the Mellin-transform of the tracer
particle mobility probability density for {\ctwo} systems:
\begin{align}
&\bar{P}[z]=\int_0^\infty \mu^{z-1} P(\mu) d\mu \nonumber\\
&=\frac{\beta}{\Gamma(\beta)^2}
\frac{B(s)^{z-1}}{\Gamma(z-1)} \frac{\Gamma(\beta
  z)^2\Gamma(\beta(z-1))\Gamma(\beta(z+1))}{\Gamma(2\beta z)}\label{eq:HTmain}
\end{align}
The inverse Mellin-transform of $\bar{P}(z)$ is an H-function
\cite{mathai}. However, due to the definition of the Mellin transform,
Eq. (\ref{eq:HTmain}) allows to directly obtain moments of the probability
distribution $P[\hat{\mu}_0(s)]$, i.e, we have 
\bee\label{eq:mob_moments}
\langle [\hat{\mu}_0(s)]^n \rangle = \bar{P}(n+1).
\eee
Unlike {\cone} systems, we note that the mobility in the {\ctwo} systems does
not self-average at long times (small Laplace frequencies), i.e., the system
does not become universal with a delta-peaked distribution of mobilities 
[compare to Eq. (\ref{eq:P_LT})]. This follows since $\langle [\hat{\mu}_0(s)]^n
\rangle$ is not simply a $n$-independent quantity to the power $n$.  Also, in
contrast to {\cone} systems, the effective medium prediction for the mean
mobility is not exact.

\section{Mean square displacement}

The results from the previous section allow us to extract the tracer
particle mean square displacement.  Employing the
fluctuation-dissipation theorem \cite{kubo66} in the form of a generalized
Einstein relation
\bee
\langle \delta x_\mathcal{T}(t)\rangle_f =\frac{F_0}{2 k_B T}\langle \delta x_\mathcal{T}^2(t)\rangle
\eee
where the subscript $f$ on the left hand side indicates that the average is performed in the presence of a constant force $f_0(t)=F_0$, whereas the average on the right hand side is in the absence of force.
Combining this with Eqs. (\ref{eq:HTmain}) and (\ref{eq:mob_moments})
for the mean mobility ($n=1$) we find the heterogeneity-averaged mean
square displacement for {\ctwo} systems:
\bee\label{eq:msd_exact} \langle
\delta x_\mathcal{T}^2(t)\rangle= \Delta (\beta) \frac{k_B T}{(\kappa \chi)^{1/2}}\frac{t^{1-\beta}}{\Gamma(2-\beta)}.
\eee 
where we have introduced a correction factor compared to the effective medium result obtained in \cite{em_sfd}:
\bee\label{eq:Delta} \Delta(\beta)=Q(\beta)
\frac{\beta}{\Gamma(\beta)}\frac{[\Gamma(2\beta)]^2\Gamma(3\beta)}{\Gamma(4\beta)}
\eee 
The inset in Fig. \ref{fig:msd} display the quantity $\Delta(\beta)$ for
the full range of $\beta$-values. We notice that the effective medium approximation gives the correct exponent $1-\beta=\alpha/(1+\alpha)$ for the heterogeneity averaged case, while the corresponding prefactor is not exact.

\section{Simulations}\label{sec:sim}

\begin{figure}[tb]
\begin{center}
\includegraphics[width=0.9\columnwidth]{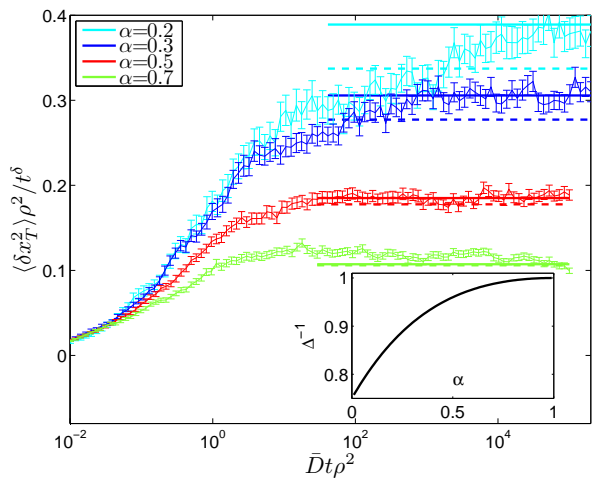}
\end{center}
\caption{Tracer particle mean square displacement for the heterogeneity-averaged case: Comparison of simulations with the effective medium (dashed lines) and
  exact (solid lines) long-time results, Eq. (\ref{eq:msd_exact}). The simulation results are averaged over 2400 realizations with the center particle taken as the tracer particle. The system size is $L=10001$ with $N=1001$ particles. The rest of the parameters are as given in Sec. \ref{sec:sim}.
(Inset) The ratio $\Delta^{-1}$, see Eq. (\ref{eq:Delta}), of
  the effective medium and exact results for the mean squared
  displacement of a tracer particle in a {\ctwo} single-file system as
  a function of $\alpha$, see Eq. (\ref{eq:prho}). Simulation data for
  $\alpha=0.3,0.5$ and $0.7$ from \cite{em_sfd}.}\label{fig:msd}
\end{figure}

In this section we provide simulation results in order to numerically test the
analytic prediction from the previous two sections.  

The simulation scheme employed here is identical to the one described in
Appendix G of \cite{em_sfd}. Briefly, each particle is placed randomly on a
line of length $L$. The particles make random jumps with a rate $q_n=2 k_B
T/(\xi_n a^2)$ and distance $l$ according to the Gaussian distribution
$P(l)=(2\pi a^2)^{-1/2}\exp[-(l-\mu)^2/(2 a^2)]$. In our simulations we use
$a=1$. The average is set to $\mu=0$ for all particles except the tagged
particle when an oscillating force is applied to it. In this case
$\mu=[F_0/(\xi_0 q_0)] \cos(\omega_0 t)$ for a force $f_n(t)=\delta_{n,0}F_0
\cos(\omega_0 t)$. The particles are hardcore interacting with a size of $b$
taken to be unity in the simulations; if an attempted jump would lead to two
particles overlapping or crossing, then the jump is either canceled or both
particles are moved according to an algorithm that preserves detailed balance
(see \cite{em_sfd} for details). Any jump that would lead to the particle
moving outside the system size $L$ is canceled. The distribution of friction
constants is taken as $\varrho(\xi_n)=A\xi_n^{-1-\alpha}$ for $\xi_n\ge\xi_c$,
$\xi_c=(A/\alpha)^{1/\alpha}$ and zero otherwise, with $A$ chosen such that
the average diffusion constant ${\bar D}=\langle k_B T/\xi_n\rangle=\alpha k_B
T/[(1+\alpha)\xi_c]$ is unity.

Let us first consider results for the tracer particle MSD.  In
Fig. \ref{fig:msd} we show comparison of simulations with the analytical
prediction for {\ctwo} systems (solid lines), Eq. (\ref{eq:msd_exact}),
showing satisfactory agreement and improving previous effective medium
predictions (dashed lines).  The correction-factor for {\ctwo} systems
$\Delta(\beta)$ is shown in Fig. \ref{fig:msd} (inset) as a function of
friction-constant exponent $\alpha$. We see that the effective medium
prediction becomes exact as $\alpha$ approaches 1, and deviates at maximum by
25\% in the limit $\alpha\to 0$. For {\cone} systems the
fluctuation-dissipation theorem combined with Eq. (\ref{eq:P_LT}) proves that
the effective medium prediction for the MSD in \cite{em_sfd} is exact for such
systems.

\begin{figure}[tb]
\begin{center}
\includegraphics[width=0.9\columnwidth]{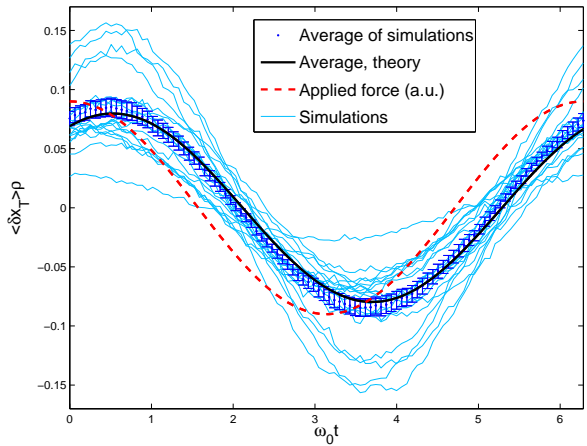}
\end{center}
\caption{Simulation results for the mean displacement from
 non-averaged simulations for 20 {\ctwo} systems with $\alpha=0.5$. Also shown are the
  average of the 20 simulations compared to the result obtained from
  Eq. (\ref{eq:HTmain}).  We used 501 particles in a box of length 5001 and an oscillation frequency
  $\omega_0/(2\pi)=10^{-5}$ and amplitude $F_0/(\xi_0 q_0)=0.002$.
\label{fig:wrapped_top}
}
\end{figure}
\begin{figure}[tb]
\begin{center}
\includegraphics[width=0.93\columnwidth]{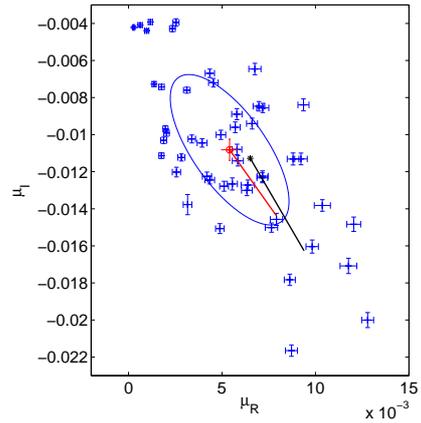}
\end{center}
\caption{Scatter plot in the complex plane (real axis placed horizontally) of
  mobilities extracted from 50 non-averaged simulations (see also Fig. \ref{fig:wrapped_top}, where results for 20 of them are shown). The circle (with errorbars) indicates the mean
  of the extracted mobilities, whereas the asterisk indicates the mean as
  obtained from Eqs. (\ref{eq:HTmain}) and (\ref{eq:mob_moments}) with $n=1$ and $s=-i\omega_0$. The
  ellipse represents the estimated covariance matrix, with the semi-major and
  semi-minor axes having lengths equal to the square root of the eigenvalues
  and pointing along the corresponding eigenvectors. The red line along the
  major axis has length equal to the square root of the difference of the
  eigenvalues. The black line line emanating from the asterisk is the analytic result for this line as obtained in Appendix \ref{appB}.
\label{fig:wrapped_bottom}
}
\end{figure}

In Fig. \ref{fig:wrapped_top} we display simulations for the mean displacement
in the presence of an oscillating force on the tagged particle
$f_0(t)=F_0\cos(\omega_0 t)$, for {\ctwo} systems. Due to the non-universality
of {\ctwo} systems each realization of friction constants gives a different
amplitude and phase for the oscillations around the mean position. The mean
value as obtained with the average mobility ($n=1$) from
Eqs. (\ref{eq:HTmain}) and (\ref{eq:mob_moments}) shows satisfactory
agreement with the simulations. Note the phase shift, $(1-\beta)\pi/2$ between
the applied force and the induced response in terms of the mean position. The
mobility as extracted from simulations is a complex quantity with real part
$\mu_R$ and imaginary part $\mu_I$. The theoretical prediction for the mean of
$\mu_R$ and $\mu_I$ are obtained from Eqs. (\ref{eq:HTmain}) and (\ref{eq:mob_moments}) by setting $n=1$
and making a Wick rotation $s=-i\omega_0$.  To assess the variability around
the mean mobility Fig. \ref{fig:wrapped_bottom} displays a scatter plot of
complex valued mobilities $\mu_{0,m}(\omega_0)$, $m=1,\dots,M$, as extracted
from simulations of $M=50$ different sets of frictions by
 \bee
\mu_{0,m}(\omega_0)=-\frac{2i\omega_0}{F_0 N}\sum_{r=1}^N e^{i\omega_0
  t_r}\llangle \delta x_{T,m}(t_r)\rrangle_{\rm sim}
\eee 
where $\delta x_{T,m}$ are deviations from the average position for friction constant set $m$, the $t_r$ runs over equally spaced times within one period of
the force and $<\dots >_{\rm sim}$ represents an average over different periods within one set of friction constants. The average squared mobility estimated over the 50 different sets of frictions is $\left<\mu_0(\omega_0)^2\right>_{\rm sim}=[-(9.4\pm 1.7)-(13.4\pm 2.4)i]\times 10^{-5}$, in satisfactory agreement with the corresponding analytic result $\left<\mu_0(\omega_0)^2\right>=(-10.1-17.6i)\times 10^{-5}$ obtained from Eqs. (\ref{eq:HTmain}) and (\ref{eq:mob_moments}) with $n=2$ and $s=-i\omega_0$.

% The theoretical results for the variations around the mean
%mobility is obtained from Eq. (\ref{eq:P_bar_final}) with $z=3$ and
%$s=-i\omega$ and displayed in the form of an ellips with minor and
%major axes determined by the eigenvalues and eigenvectors of the
%covariance matrix for the fluctuations of $\mu_R$ and $\mu_I$.  The
%direction of the major axis is proportional a vector with components
%$\llangle \mu_R\rrangle $ and $\llangle \mu_I\rrangle$, see the
%Supplemental Material for details. We notice the nice agreement
%with theory with $X \%$ of simulation data points within the
%theoretically predicted ellips.
 
\section{Conclusion}

Exactly solvable many-body models have over the years served all
fields of physics, chemistry and biological sciences. The present
manuscript provides important insights into the combined effects of
heterogeneity and particle-particle interactions on the dynamics in
stochastic processes. % In particular, we have found that the mobility of a particle behaves universally in systems with finite average firction, in contrast to systems with heavy-tail distributed frictions.
In particular, we provided an asymptotically exact analytic
expression for the probability density of mobility in a
single-file system, and for harmonically coupled beads, with different
frictions constants. 
Our study paves the way for force response
studies of other complex heterogeneous many-body systems.

We hope that the type of system introduced here will find experimental
realizations for transport processes where heterogeneity is prominent -
examples include, motion of flourescently labeled proteins on DNA molecules
and other macromolecules, or diffusion of dissimilar particles in
nanochannels.

\section{Acknowledgements}

T.A. acknowledges funding from the Knut \& Alice Wallenberg Foundation
and the Swedish Research Council (grant no. 2009-2924). Computer time
was provided by the Danish Center for Scientific Computing.

\appendix

\section{Asymptotic solution of Eq. (\ref{eq:g_s_eq})}\label{appA}

Let us consider the expression for the distribution $g_s(m)$, Eq. (\ref{eq:g_s_eq})
in the main text, for the case of long times ($s\rightarrow 0$). In this limit
Eq. (\ref{eq:g_s_eq}) can be solved following the approach in the appendix of
Ref. \cite{bernasconi_80}. However, as this derivation is lengthy we here
provide a simpler as well as more general version of the derivation. Following
Ref. \cite{bernasconi_80} we write
\bee\label{eq:g_w_scaling}
g_s(m)=\frac{1}{\epsilon(s)} h\big{(}
m/\epsilon(s)\big{)}
\eee
with the $s$-dependent scaling function $\epsilon=\epsilon(s)$
chosen to be positive and satisfying $\epsilon(s)\rightarrow 0$ and $s/\epsilon(s)\to 0$ as
$s\rightarrow 0$. Taking the Mellin-transform
[$\bar{A}(z)=\int_0^\infty x^{z-1} A(x) dx $] of Eq. (\ref{eq:g_s_eq})
with respect to $m$ and substituting the integration variable with $v=m'/\epsilon$ we get:
\bee\label{eq:starting_eq}
\bar{h}(z)=\int_0^\infty dv\, h(v) \int_0^\infty dy\, 
R(y) \left( \frac{\epsilon}{y}+\frac{1}{v+s/(\kappa \epsilon) } \right)^{1-z}
\eee
which is the starting point of our simplified derivation.

\subsection{{\cone} systems}

Let us first consider light-tailed (LT) systems. For these we expand the
right-hand-side of Eq. (\ref{eq:starting_eq}) to first subleading order in
$\epsilon$ and $s/(\kappa \epsilon)$. Also making use of the definition of a
Mellin-transform we get that the right-hand side equals $\bar{h}(z)-(1-z)(s/\kappa
\epsilon) \bar{h}(z-1)+(1-z)\langle y^{-1} \rangle \epsilon
\bar{h}(z+1)$. Eq. (\ref{eq:starting_eq}) then becomes
\bee\label{eq:LT_Mellin}
\frac{s}{\kappa\bar{\xi} \epsilon^2} \bar{h}(z-1)= \bar{h}(z+1)
\eee
We obtain a non-trivial solution for $\bar{h}$ by choosing 
\bee\label{eq:eps_C1}
\epsilon(s)=\left(\frac{s}{\kappa \bar{\xi}}\right)^{1/2}
\eee
where $\bar{\xi}=\int_0^\infty (1/y) R(y) dy = \int_0^\infty \xi \prho(\xi)
d\xi$ is the mean friction constant.  The solution to Eq. (\ref{eq:LT_Mellin})
with $\bar{h}(1)=1$ (normalization condition) is simply $ \bar{h}(z)=1$ which
when Mellin-inverted gives
\bee\label{eq:h_C1}
h(x)=\delta(x-1)
\eee
in agreement with \cite{bernasconi_80}.

\subsection{{\ctwo} systems}

For heavy-tailed (HT) systems the analysis is slightly more involved. 
First we integrate Eq. (\ref{eq:starting_eq}) by parts,
introducing the cumulative distribution $C(y)=\int_0^y
R(y')dy'$, to find $\bar{h}(z)=T_1+T_2$ with
\begin{align}
T_1&=\int_0^\infty dv\, h(v)  C(y)
\left.\left[\frac{\epsilon}{y} +\frac{1}{v+s/(\kappa \epsilon)}\right]^{1-z}\right|_{y=0}^\infty\\
T_2&=\int_0^\infty dv\, h(v) \int_0^\infty
dy\frac{(1-z)\epsilon}{y^2} C(y) \left[\frac{\epsilon}{y}+\frac{1}{v+s/(\kappa\epsilon)}\right]^{-z}
\end{align}
Since $R(y)\sim A/y^{1-\alpha}$ for small $y$ we have $C(y)\sim A y^{\alpha}/\alpha$. Restricting $z$ to $z>1-\alpha$ the lower boundary term above vanishes. At the opposite boundary we have $C(\infty)=1$ and thus we get
\bee
T_1=\int_0^\infty dv h(v)v^{z-1} \left[1+\frac{s}{\kappa \epsilon v}\right]^{z-1}
\eee
For $T_2$, if we set $y=\epsilon y'$ and let $\epsilon\to 0$ we have $C(\epsilon y')\sim A(\epsilon y')^\alpha/\alpha$ and find to leading order in $\epsilon$ and $s/\epsilon$:
\begin{align}
T_2&\sim 
\int_0^\infty dv\, h(v) \int_0^\infty
dy'\frac{(1-z)}{ (y')^2} \frac{A(\epsilon y')^\alpha}{\alpha} \left[\frac{1}{y'}+\frac{1}{v}\right]^{-z}\nonumber\\
&=\epsilon^{\alpha}\frac{A(1-z)\Gamma(1-\alpha)\Gamma(z+\alpha-1)}{\alpha \Gamma(z)} {\bar{h}}(z+\alpha)\label{eq:T2}
\end{align}
%(1-z)\frac{\Gamma(z-1+\alpha)\Gamma(1-\alpha)}{\Gamma(z)} \epsilon^\alpha y_c^{-\alpha}\nonumber\\
%\times \int_0^\infty dv
%h(v) (1+\frac{s}{\kappa \epsilon v})^{z-c}
%\eea
%
Similarly we expand $T_1$, but here we include the subleading term in $s/(\kappa \epsilon)$
\bee
T_1\sim \bar{h}(z)+\frac{(z-1)s}{\kappa \epsilon}\bar{h}(z-1)\label{eq:T1}
\eee
From Eqs. (\ref{eq:T2}) and (\ref{eq:T1}) we see that a way to obtain a non-trivial equation for $\bar{h}(z)$ in the limit $s\to 0$ is to choose
\bee\label{eq:eps_C2}
\epsilon(s)=\left(\frac{\alpha s}{\kappa A}\right)^\beta
\eee
with $\beta=1/(1+\alpha)$. With this choice we find that $\bar{h}(z)$ satisfies
\bee\label{eq:HT_Mellin}
\bar{h}(z-1)=
\frac{\Gamma(z+\alpha-1)\Gamma(1-\alpha)}{\Gamma(z)}\bar{h}(z+\alpha)
\eee
which has the solution (with $\bar{h}(1)=1$)
\bee\label{eq:h_bernasconi}
\bar{h}(z)=\frac{\beta}{\Gamma(\beta)}[\beta^2 \Gamma(1-\alpha)]^{\beta(1-z)}
\frac{\Gamma(\beta(z-1))\Gamma(\beta z)}{\Gamma(z-1)}
\eee
in agreement with \cite{bernasconi_80}.

\section{The covariance matrix}\label{appB}

Writing the mobility in Fourier space according to (with $\delta\mu_R$ and $\delta\mu_I$ real)
\begin{equation}
\mu_0(\omega_0)=\left<\mu_0(\omega_0)\right>+\delta\mu_R+i\delta\mu_I
\end{equation}
we can define a covariance matrix for the real and imaginary deviations
\begin{equation}
\Sigma=\left(\begin{array}{c c}\left<\delta\mu_R^2\right> & \left<\delta\mu_R\delta\mu_I\right>\\ \left<\delta\mu_R\delta\mu_I\right> & \left<\delta\mu_I^2\right> \end{array}\right)
\end{equation}
From simulation data with estimates $\mu_{0,m}(\omega_0)$, $m=1,\dots,M$, of mobility from $M$ different sets of friction coefficients we estimate the average as $\left<\mu_0(\omega_0)\right>_{\rm est}=\sum_{m=1}^M \mu_{0,m}(\omega_0)/M$. For the deviations $\delta\mu_{R,m}+i\delta\mu_{I,m}=\mu_{0,m}(\omega_0)-\left<\mu_0(\omega_0)\right>_{\rm est}$ we estimate the components of $\Sigma$ by
\begin{equation}
\left<\delta\mu_A\delta\mu_B\right>_{\rm est}= \frac{1}{M-1}\sum_{m=1}^M \delta\mu_{A,m}\delta\mu_{B,m}
\end{equation}
where $A$ and $B$ are either of $R$ or $I$. The ellipse in Fig. 3 has been drawn with its center on $\left<\mu_0(\omega_0)\right>_{\rm est}$, and with the major and minor radii equal to the square roots of the eigenvalues of the estimated $\Sigma$. The corresponding axes points along the eigenvectors.

The analytic result for the second moment about the mean: $\left<\left(\delta\mu_R+i\delta\mu_I\right)^2\right>$ allows us to extract two results regarding the covariance matrix $\Sigma$. One regards its eigenvalues $\lambda_\pm$
\begin{equation} 
\lambda_\pm=\frac{\left<\delta\mu_R^2\right>+\left<\delta\mu_I^2\right>\pm\sqrt{(\left<\delta\mu_R^2\right>-\left<\delta\mu_I^2\right>)^2+4\left<\delta\mu_R\delta\mu_I\right>^2}}{2}
\end{equation}
The difference of these can be found from the analytic calculation of the moments as
\begin{align}
\label{eq:difference_lambda}
\lambda_+-\lambda_-&=\sqrt{\left(\left<\delta\mu_R^2\right>-\left<\delta\mu_I^2\right>\right)^2+4\left<\delta\mu_R\delta\mu_I\right>^2}\nonumber\\
&=\left|\left<\left(\delta\mu_R+i\delta\mu_I\right)^2\right>\right|\nonumber\\
&=\left|\left<\mu_0(\omega_0)^2\right>-\left<\mu_0(\omega_0)\right>^2\right|
\end{align}
Note that Eq. (\ref{eq:difference_lambda}) implies that $\lambda_+\ge
\lambda_-$. 
We can also extract the eigenvectors. To see this first note that the
asymptotic result $\left<\mu_0(\omega_0)^n\right>\simeq (-i\omega_0)^{n\beta}$
for the $n$th moment tells us that
$\left<\left(\delta\mu_R+i\delta\mu_I\right)^2\right>=Ce^{2i\phi}$, where
$C>0$ and
$e^{i\phi}=\left<\mu_0(\omega_0)\right>/|\left<\mu_0(\omega_0)\right>|$, i.e.,
the phase of the second moment is twice that of the first moment. Using this
we find that the eigenvector corresponding to $\lambda_+$ is:
\begin{align}
\vec{v}_+&=\left(\begin{array}{c}\left<\delta\mu_R\delta\mu_I\right>\\ \lambda_+-\left<\delta\mu_R^2\right>\end{array}\right)=\frac{C}{2}\left(\begin{array}{c}\sin 2\phi\\ 1-\cos 2\phi\end{array}\right)\nonumber\\
&=C\sin\phi\left(\begin{array}{c}\cos \phi\\ \sin\phi\end{array}\right)=\frac{C\sin\phi}{|\left<\mu_0(\omega_0)\right>|}\left(\begin{array}{c}{\rm Re} \left<\mu_0(\omega_0)\right>\\ {\rm Im}\left<\mu_0(\omega_0)\right>\end{array}\right)
\end{align}
Thus, for the ellipse plotted in Fig. 3 of the main text we find that the major axis should point along the line to the origin. The black line in Fig. 3 drawn from the asterisk has length $\sqrt{\lambda_+-\lambda_-}$ and points along $-\vec{v}_+$ as obtained from the analytic results.

\bibliographystyle{unsrt}
\bibliography{refs}

\begin{thebibliography}{10}

\bibitem{kubo_1986}
R. Kubo, Science {\bf 233}, 330 (1986).

\bibitem{marconi_2008}
U.M.B. Marconi, A. Puglisi, L. Rondini, A. Vulpiani, Phys. Rep. {\bf 461}, 111
  (2008).

\bibitem{bustamante_04}
C. Bustamante, Y.R. Chemla. N.R. Forde and D. Izhaky, Ann. Rev. Biochem. {\bf
  73}, 705 (2004).

\bibitem{jarzynski}
C. Jarzynski, Phys. Rev. Lett. {\bf 78}, 2690 (1997).

\bibitem{gorissen2012}
M. Gorissen, Alexandre Lazarescu, K. Mallick and C. Vanderzande, Phys. Rev.
  Lett. {\bf 109}, 170601 (2012).

\bibitem{HA}
T.~E.~Harris, J.~Appl.~Prob. {\bf 2}(2), 323 (1965).

\bibitem{Kollmann_03}
M. Kollmann, Phys. Rev. Lett. {\bf 90}, 180602 (2003).

\bibitem{Taloni_06}
F. Marchesoni and A. Taloni, Phys. Rev. Lett. {\bf 97}, 106101 (2006).

\bibitem{barkai_09}
E. Barkai and R. Silbey, Phys. Rev. Lett. {\bf 102}, 050602 (2009).

\bibitem{Aslangul_00}
C. Aslangul, J. Phys. A {\bf 33}, 851 (2000).

\bibitem{Brzank2}
A. Brzank and G.M. Sch\"utz, J. Stat. Mech: Theory and Experiment P08028
  (2007).

\bibitem{JALA}
P. Goncalves and M. D. Jara, J. Stat. Phys. {\bf 132}, 1135 (2008).

\bibitem{Jara_09}
M. Jara, e-print: arXiv:0901.0229.

\bibitem{TA_etal}
T. Ambj\"ornsson, L. Lizana, M.A. Lomholt and R.J. Silbey, J. Chem. Phys. {\bf
  129}, 185106 (2008).

\bibitem{Flomenbom}
O. Flomenbom, Phys. Rev. E {\bf 82}, 031126 (2010).

\bibitem{em_sfd}
M.A. Lomholt, L. Lizana and T. Ambj\"ornsson, J. Chem. Phys. {\bf 134}, 045101
  (2011).

\bibitem{KKDGPRSUK}
V. Kukla, J. Kornatowski, D. Demuth, I. Girnus, H. Pfeifer, L. Rees, S. Schunk,
  K. Unger, and J. K\"arger, Science {\bf 272}, 702 (1996).

\bibitem{meersmann00}
T. Meersmann, J.W. Logan, R. Simonutti, S. Caldarelli, A. Comotti, P. Sozzani,
  L.G. Kaiser and A. Pines, J. Phys. Chem. A {\bf 104}, 11665 (2000).

\bibitem{hahn96}
K. Hahn, J. K{\"a}rger, and V. Kukla, Phys. Rev. Lett. {\bf 76}, 2762 (1996).

\bibitem{WBL}
Q.~H.~Wei, C.~Bechinger, P.~Leiderer, Science {\bf 287}, 625 (2000).

\bibitem{gupta95}
V. Gupta, S.S. Nivarthi, A.V. McCormick, H. Ted Davis, Chem. Phys. Lett. {\bf
  247}, 596 (1995).

\bibitem{HOKE}
A. L. Hodgkin and R. D. Keynes, J. Physiol. (London) {\bf 128}, 61 (1955).

\bibitem{vermeulen_2002}
K.C. Vermeulen, G.J. Stienen, C.F. Schmid, J Muscle Res Cell Motil. {\bf 23} 71
  (2002).

\bibitem{berger_2012}
F. Berger, C. Keller, S. Klumpp, and R. Lipowsky, Phys. Rev. Lett. {\bf 108},
  208101 (2012).

\bibitem{Elf_09}
G.-W. Li, O.G. Berg and J. Elf, Nature Phys. {\bf 5}, 294 (2009).

\bibitem{kubo66}
R. Kubo, Rep. Prog. Phys. {\bf 29}, 255 (1966).

\bibitem{lizana_10}
L. Lizana, T. Ambj\"ornsson, A. Taloni, E. Barkai, M. A. Lomholt, Phys. Rev. E
  {\bf 81}, 051118 (2010).

\bibitem{footnote1}
The analysis performed here is similar to the one in appendix A of Ref.
  \cite{em_sfd} (in Laplace-space rather than in the Fourier domain). The
  relation between our $m_n$'s and the quantities in Ref. \cite{em_sfd} is:
  $m_n^{(\pm)}(s)= [\xi_n +\gamma_n^{(\pm)}(s)]^{-1}$.

\bibitem{bernasconi_80}
J. Bernasconi, W.R. Wyss and W. Wyss, Z. Physik B {\bf 37}, 175 (1980).

\bibitem{alexander81}
S. Alexander, J. Bernasconi, and W. R. Schneider, Rev. Mod. Phys. {\bf 53}, 175
  (1981).

\bibitem{oberhettinger}
F. Oberhettinger, Tables of Mellin Transforms (Springer) New York (1974).

\bibitem{ABST}
Milton Abramowitz and Irene A. Stegun, {\it Handbook of Mathematical Functions
  with Formulas, Graphs, and Mathematical Tables}, (Dover, New York, 1964).

\bibitem{mathai}
A.M. Mathai, R.K. Saxena and H.J Haubold, The H-Function: Theory and
  Applications (Springer) 2010.

\end{thebibliography}

\end{document}